\documentclass{article}
\usepackage{arxiv}
\usepackage[english]{babel}
\usepackage[utf8]{inputenc}
\usepackage[T1]{fontenc}
\usepackage{amsmath,amsfonts,amssymb}
\usepackage{graphicx}
\usepackage{array,tabularx,multirow}
\usepackage[table]{xcolor}
\usepackage{cite}
\usepackage{authblk}
\usepackage{placeins}
\usepackage[font=small,justification=raggedright,singlelinecheck=false,hypcap=false]{caption}
\usepackage{wrapfig}
\usepackage{needspace}
\usepackage[colorlinks=true,allcolors=blue]{hyperref}
\newcolumntype{Y}{>{\raggedright\arraybackslash}X}
\newcommand{\shorttitle}{Trustworthy AI and Democracy}
\renewcommand{\headeright}{Author manuscript}

\title{Trustworthy AI and Democracy: A Dual Taxonomy and Operational Framework of Democratic Risks and Contributions}
\author[1]{Oier Mentxaka}
\author[1]{Natalia Díaz-Rodríguez}
\author[2]{Mark Coeckelbergh}
\author[3,4,5]{Marcos López de Prado}
\author[6]{Emilia Gómez}
\author[6,7]{David Fernández-Llorca}
\author[1]{Enrique Herrera-Viedma}
\author[1,5]{Francisco Herrera}
\affil[1]{Department of Computer Science and Artificial Intelligence, DaSCI, University of Granada, Granada, Spain}
\affil[2]{Department of Philosophy, University of Vienna, Vienna, Austria}
\affil[3]{School of Engineering, Cornell University, Ithaca, NY, USA}
\affil[4]{Department of Mathematics, Khalifa University of Science and Technology, Abu Dhabi, United Arab Emirates}
\affil[5]{ADIA Lab, Al Maryah Island, Abu Dhabi, United Arab Emirates}
\affil[6]{European Commission, Joint Research Centre (JRC), Seville, Spain}
\affil[7]{Computer Engineering Department, University of Alcalá, Alcalá de Henares, Spain}
\date{}
\begin{document}
\maketitle
\begin{center}
\small Corresponding author: Oier Mentxaka (\href{mailto:oiermentxaka@ugr.es}{oiermentxaka@ugr.es}).\\[4pt]
Published in \emph{IEEE Transactions on Systems, Man, and Cybernetics: Systems} (2026).\\
DOI: \href{https://doi.org/10.1109/TSMC.2026.3743429}{10.1109/TSMC.2026.3743429}
\end{center}
\medskip

\begin{abstract}
Artificial Intelligence (AI) can both undermine and strengthen democratic governance. This paper introduces the AIRD/AIPD framework, a dual taxonomy and operational framework for analyzing AI’s democratic risks and conditional contributions. The \textit{AI Risks to Democracy (AIRD)} taxonomy identifies threats to autonomy, fairness, public discourse, democratic institutions, and trust, while the \textit{AI’s Positive Contributions to Democracy (AIPD)} taxonomy captures AI’s potential to enhance efficiency, accessibility, security, informed citizenry, transparency and accountability, inclusive democratic dialogue, and evidence-based policymaking.

Grounded in five democratic principles and in the seven Trustworthy AI requirements proposed by the European Commission’s High-Level Expert Group on AI, the framework connects democratic goals with governance and design mechanisms. Beyond classification, it provides an operational process for identifying affected contexts and actors, classifying primary and secondary AIRD/AIPD effects, mapping them to relevant Trustworthy AI requirements, and defining proportionate governance responses and responsible actors. A compact operational mapping and two worked examples illustrate its application to electoral misinformation and AI-assisted public service allocation.

The analysis highlights the transversal relevance of transparency, accountability, fairness, and societal wellbeing, while recognizing trade-offs, implementation costs, differential impacts, institutional capacity, and contextual variation across democratic settings. The framework is conceptual and has not yet been empirically validated, but it offers a traceable and adaptable basis for research, policy analysis, institutional oversight, and responsible AI design. By connecting democratic diagnosis with governance action, it helps bridge the gap between normative principles and the operational requirements of trustworthy and democratically legitimate AI.
\end{abstract}


\section{Introduction}

The accelerated deployment of artificial intelligence (AI) in multiple sectors has begun to shape not only technological infrastructures, but also the social, economic, and political foundations of democratic societies. As AI systems increasingly mediate communication, decision-making, and policy implementation, they raise significant concerns regarding the protection of democratic values such as freedom, equality, fraternity, the rule of law, and tolerance. Although many studies have explored the ethical implications of AI \cite{floridi2019unified, fjeld2020principled, radanliev2025ai}, fewer have fully examined its democratic implications, both in terms of the risks it poses and the potential it holds to strengthen democratic institutions \cite{coeckelbergh2025llms, saura2025synthetification, calvo2025generative, summerfield2025impact}. This ambivalent relationship has recently been described as a democratic dilemma: AI can both expand democratic capacities and intensify vulnerabilities in representation, accountability, and public trust \cite{altman2026democracy}.

The need for such a democracy-specific framework has become even clearer in recent research. Wiewiura et al. \cite{wiewiura2026ai} show that scholarship on AI and democracy often relies on different, and sometimes implicit, conceptions of democracy, which can lead empirical and normative debates to talk past one another. Rymon \cite{rymon2026representatives} further demonstrates that AI transforms democratic representation by reshaping both input representation, through preference aggregation and deliberative platforms, and output representation, through algorithmic decision-making systems. Schroeder et al. \cite{schroeder2026malicious} identify malicious AI swarms as an emerging threat to democratic information ecosystems, while Kalsnes and Schia \cite{kalsnes2026protecting} show that early codes of conduct for generative AI use during elections often lack enforceability, consistency, and normative clarity. Chehoudi \cite{chehoudi2026artificial}, in turn, provides comparative evidence suggesting that the relationship between AI development and democratic quality depends on political and institutional conditions. Taken together, these recent contributions reinforce the need for a structured taxonomy that clarifies democratic concepts, distinguishes risks from positive contributions, and connects both to concrete governance mechanisms.

To provide a structured understanding of this duality, this paper introduces a framework grounded in normative democratic theory and applied AI ethics. Our analysis draws on epistemic democracy, which emphasizes the capacity of democratic procedures to support better collective judgment \cite{estlund2008democratic, goodin2018epistemic, landemore2012democratic}; deliberative and participatory theories, which stress public reasoning, inclusive participation, and discursive legitimacy \cite{fishkin1991democracy, habermas1996between, dryzek2000deliberative}; and agonistic and critical approaches, which foreground conflict, pluralism, and power \cite{mouffe2000democratic}. The five democratic principles used in this paper should therefore be read as a synthesis of these strands rather than as a novel theory of democracy. The analysis is then connected to the seven requirements for Trustworthy AI proposed by the European Commission’s High-Level Expert Group on AI \cite{hleg2019ethics}, which provide a practical reference for evaluating and governing AI systems.

This study addresses the need for a shared vocabulary and a governance-oriented framework to assess AI's democratic implications. Despite growing attention to AI regulation, public and scholarly debates often discuss technical or ethical issues separately from their democratic effects. This leaves a gap in policymaking, research, and civic understanding: the lack of a structured framework connecting AI-related risks and opportunities to democratic values, Trustworthy AI requirements, and governance responses.

To fill this gap, we propose the AIRD/AIPD framework. The AI Risks to Democracy (AIRD) taxonomy classifies how AI systems, when misaligned with ethical and democratic principles, may threaten autonomy, fairness, power relations, public discourse, democratic institutions, and trust. The AI Positive Contributions to Democracy (AIPD) taxonomy identifies the conditional ways in which responsibly governed AI may support democratic processes through efficiency, accessibility, security, informed citizenry, transparency and accountability, global and inclusive democratic dialogue, and evidence-based policymaking. Compared with broader democratic principles or general Trustworthy AI frameworks, the AIRD/AIPD framework adds a democracy-specific analytical layer by translating abstract values into categories of AI-related risks and opportunities linked to concrete uses, affected democratic processes, and governance responses.


The distinctive contribution of this paper is a democracy-specific framework linking AI-related risks and conditional contributions to democratic principles, Trustworthy AI requirements, governance responses, and responsible actors. Building on approaches that connect AI principles to responsible systems and regulation~\cite{diaz2023connecting}, AIRD/AIPD identifies the affected democratic process and makes the rationale for each governance response explicit. Its dual taxonomy accommodates overlapping risks and opportunities, while its operational workflow distinguishes risk mitigation from the conditions needed to realize democratic benefits. These mappings support structured assessment; they do not establish the causal effectiveness of the proposed measures.

The central hypothesis of this paper is that the dual nature of AI, as both a threat and a support for democracy, can be systematically understood, analyzed, and addressed through structured taxonomies aligned with ethical and policy-based AI governance principles. In this sense, the AIRD/AIPD framework should be understood as an analytical structure rather than as a rigid typology. AI-related phenomena may affect democracy through several mechanisms at once, so the categories are not mutually exclusive: a given case may be assigned a primary label, corresponding to its dominant democratic risk or opportunity, while secondary labels can capture additional effects.

The development of the AIRD/AIPD framework followed a conceptual and thematic synthesis of recent scholarship on AI, democracy, and Trustworthy AI governance. The literature was identified through targeted searches in major academic databases and through backward and forward reference tracking of key works on AI ethics, democratic theory, algorithmic governance, electoral integrity, AI-mediated deliberation, and public-sector AI. We prioritized peer-reviewed journal articles, book chapters, institutional reports, and highly relevant preprints published since 2021, with additional attention to closely related works published during the revision period. We included publications that explicitly addressed the relationship between AI systems and democratic values, institutions, participation, representation, accountability, public trust, or governance, and excluded from the core classification works focused solely on technical performance unless they were needed to clarify relevant technical background.

The thematic classification proceeded in three steps. First, we extracted recurring democratic risks and positive contributions discussed in the literature. Second, we grouped these recurring themes according to their dominant democratic implication, distinguishing AIRD from AIPD. Third, we mapped the resulting categories onto the five democratic principles and the seven Trustworthy AI requirements used in the paper. Because AI-related phenomena often have overlapping effects, the taxonomy allows multi-label classification: a given case may instantiate a primary AIRD or AIPD category while also presenting secondary effects. The framework is therefore intended as a structured conceptual mapping rather than a systematic review or an empirically validated causal model; its aim is not to eliminate interpretive judgment, but to make the basis for classification explicit and traceable.
Only publications directly addressing the democracy--AI nexus informed the core classification. In line with the legal scope of the EU AI Act, our analysis focuses on civilian applications of AI and their implications for democratic institutions and practices. We do not consider defense or military uses of AI,\footnote{The EU AI Act excludes AI systems used exclusively for military, defence, or national security purposes from its material scope. Complementary guidance on trustworthy AI in the defense domain is being developed separately, for example in the European Defence Agency White Paper \emph{Trustworthiness for AI in Defence} \cite{eda2025trustworthiness}.} which fall under distinct regulatory arrangements and normative questions.

Grounded in the EU Trustworthy AI framework and liberal-democratic values, AIRD/AIPD may require adaptation to local political, legal, institutional, and cultural conditions. It is a flexible basis for contextualization, comparative research, and implementation, not a one-size-fits-all model.

Section \ref{sec:Pre} presents the conceptual foundations; Sections \ref{sec:positive} and \ref{sec:risk} introduce AIPD and AIRD; Section \ref{sec:maping} maps risks to requirements, governance responses, and responsible actors; Section \ref{sec:reflect} discusses application, contextualization, trade-offs, and limitations; and Section \ref{sec:Con} concludes.
 
\section{Preliminaries: Principles for Democracy and Trustworthy AI}
\label{sec:Pre}

This section introduces the conceptual foundations used throughout the paper: five democratic principles that serve as benchmarks for assessing AI's societal impacts, and the seven Trustworthy AI requirements proposed in the EU Ethics Guidelines \cite{hleg2019ethics}. Together, these frameworks provide the normative and operational basis for analyzing both the democratic risks and contributions of AI.

\Needspace{0.60\textheight}
\subsection{Democratic Principles} \label{sec:demoDef}

\Needspace{0.56\textheight}
\begin{wrapfigure}{r}{0.30\textwidth}
\centering
    \includegraphics[width=\linewidth]{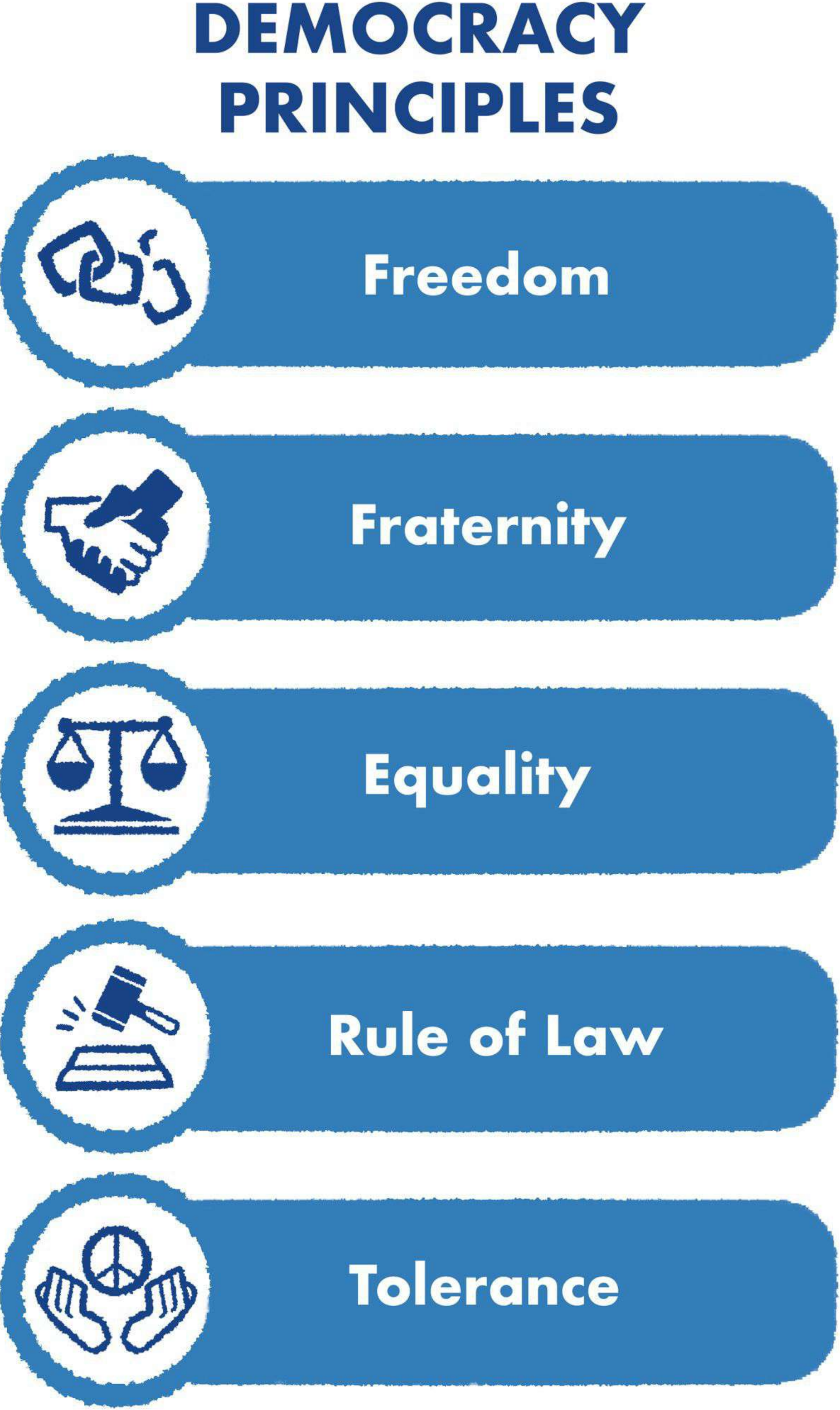}
    \caption{The five democratic principles proposed in \cite{coeckelbergh2024why}.}   
\label{fig:democracyPrinciples}
\end{wrapfigure}
Building on Coeckelbergh's political-epistemological account of democracy and AI \cite{coeckelbergh2024why} and on widely recognized strands of democratic theory, we understand democracy as a set of institutional and social arrangements that secure meaningful citizen agency in collective decision-making. Deliberative, participatory, epistemic, and agonistic approaches stress fair inclusion in public reasoning, responsible political judgment, and the need to address persistent power asymmetries \cite{fishkin1991democracy, habermas1996between, dryzek2000deliberative, estlund2008democratic, goodin2018epistemic, landemore2012democratic, mouffe2000democratic}. Across these debates, five values recur as enabling conditions for democratic agency: Freedom, Fraternity, Equality, Rule of Law, and Tolerance. In line with Coeckelbergh's analysis, we use these principles as the democratic benchmarks for assessing how AI can either support or undermine democracy, as summarized in Fig.~\ref{fig:democracyPrinciples}.

\noindent \textit{P1: Freedom}: Freedom of action, expression, belief, and thought without oppression.

\noindent \textit{P2: Fraternity}: Solidarity, belonging, mutual support, and respect for dignity; perceived unfairness can weaken fraternal relations.

\noindent \textit{P3: Equality}: Non-discriminatory access to rights and opportunities, recognizing that formally equal decisions may remain unfair.

\noindent \textit{P4: Rule of law}: Equal accountability under the law, fair process, and protection of individual rights.

\noindent \textit{P5: Tolerance}: Respect for diverse opinions, beliefs, and practices, enabling dialogue and coexistence.

\subsection{Trustworthy AI principles and requirements} \label{sec:taiDefs}

The HLEG's Ethics Guidelines for Trustworthy AI \cite{hleg2019ethics}, OECD AI Principles \cite{oecd2019principles}, NIST AI Risk Management Framework \cite{nist2021airmf}, and UNESCO Recommendation on the Ethics of AI \cite{unesco2021ethics} consolidate the notion of Trustworthy AI. Following the HLEG framework, we adopt its three pillars and seven requirements:

\begin{enumerate}
    \item \textit{Lawfulness}: AI systems shall comply with applicable legal frameworks, including the GDPR \cite{act2022GDPR}, sector-specific legislation, the AI Act \cite{act2024AIA}, and the Digital Services Act (DSA) \cite{act2022DSA}, which targets systemic risks linked to very large online platforms (VLOPs) and search engines (VLOSEs), not other intermediary services. Trustworthy AI thus requires legal compliance while also informing the evolution of legal norms.
    \item \textit{Ethics}: Beyond compliance, Trustworthy AI should uphold ethical norms, fundamental rights, and values such as fairness, justice, and human dignity \cite{estevez2022glossary}, particularly as generative AI and large language models often outpace regulatory responses.
    \item \textit{Robustness}: Trustworthy AI shall ensure safety and reliability, minimizing unintentional harm from both technical and social perspectives \cite{hendrycks2025introduction}.
\end{enumerate}

The HLEG framework further derives four ethical principles from fundamental human rights:

\begin{enumerate}
    \item \textit{Respect for human autonomy}: AI systems shall preserve freedom and autonomy, avoid manipulation, ensure human supervision, and support human-centered design.
    
    \item \textit{Prevention of harm}: AI systems shall be secure and robust, avoiding new or aggravated harms, particularly for vulnerable people and in contexts of power or information asymmetry.

    \item \textit{Fairness}: AI systems shall ensure a fair distribution of benefits and costs, avoid bias and discrimination, and allow people to object to automated decisions.
    
    \item \textit{Explainability}: AI systems shall be transparent, traceable, and explainable, especially in “black box” cases, enabling users to challenge decisions.
\end{enumerate}

\begin{wrapfigure}{r}{0.30\textwidth}
\centering
    \includegraphics[width=\linewidth]{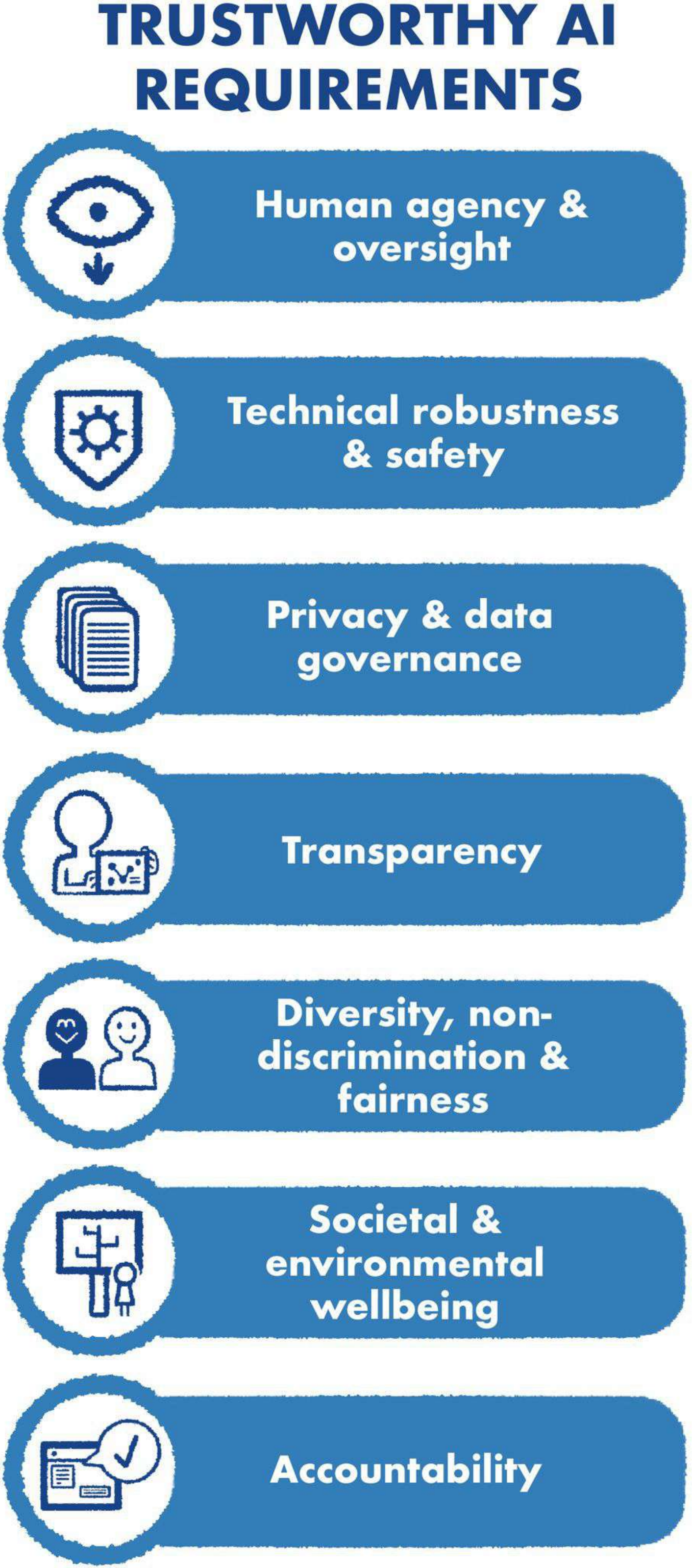}
    \caption{The seven requirements for Trustworthy AI (adapted from \cite{hleg2019ethics}).}
    \label{fig:trustworthyRequirements}
\end{wrapfigure}

Based on these four principles, seven requirements were proposed (illustrated in Figure \ref{fig:trustworthyRequirements}):

\Needspace{0.56\textheight}
\noindent \textit{Rq1. Human Agency and Oversight}: AI systems should support human agency, autonomy, and fundamental rights, ensuring meaningful human oversight.

\noindent \textit{Rq2. Technical Robustness and Safety}: AI should be resilient to attacks, secure, reliable, accurate, and reproducible, with fallback mechanisms to ensure safety.

\noindent \textit{Rq3. Privacy and Data Governance}: AI must respect privacy, ensure responsible data handling, and adhere to principles of data quality, integrity, and access control.

\noindent \textit{Rq4. Transparency}: AI should be explainable, traceable, and effectively communicated to stakeholders.

\noindent \textit{Rq5. Diversity, Non-Discrimination, and Fairness}: AI should prevent unfair bias, promote accessibility and universal design, and encourage stakeholder participation.

\noindent \textit{Rq6. Societal and Environmental Wellbeing}: AI should consider its impact on society, democracy, and the environment, promoting sustainability and social responsibility.

\noindent \textit{Rq7. Accountability}: AI should be auditable, incorporate effective risk management, and include mechanisms for responsibility assignment, impact minimization, reporting, and redress.

\noindent Transparency (Rq4) requires explanations that support scrutiny and contestation. However, explaining a model’s output does not establish how an intervention would affect democratic outcomes. Building on Herrera’s discussion of XAI~\cite{herrera2025reflections} and López de Prado’s treatment of causal inference~\cite{de2023causal}, we distinguish model interpretation from causal evaluation. Feature attributions and hypothetical changes to model inputs do not, by themselves, identify the effects of changing a policy or deployment practice. Such claims require a specified intervention, outcome, comparison, and defensible identification assumptions. Causal evidence can inform democratic judgment, but cannot determine which objectives or trade-offs are legitimate.

At the same time, prior work shows that opacity is not only technical, but also institutional, strategic, expert, and emergent. In judicial, administrative, and political contexts, opaque AI systems may reproduce bias, shield decisions from scrutiny, consolidate technocratic control, deepen power asymmetries, weaken public trust, and reduce citizens' ability to question or influence AI-mediated decisions \cite{winter2022challenges, buhmann2023deep, da2023framework, innerarity2024artificial, duberry2022artificial, beckman2024artificial}. Therefore, transparent and accountable governance mechanisms are necessary to prevent AI systems from bypassing democratic deliberation and undermining the broader aims of Trustworthy AI.

\subsection{Complementarity of Frameworks: Democracy as Goal, Trustworthy AI as Means}

Democratic principles define the normative goals and benchmarks for assessing AI's democratic harms and benefits; Trustworthy AI requirements provide practical design and governance means to safeguard them. Together, they structure the analysis of AI's capacity to undermine or reinforce democracy.

\section{Positive Taxonomy: AI’s Positive Contributions to Democracy}
\label{sec:positive}

When developed and deployed responsibly, AI can offer meaningful opportunities to strengthen democratic institutions. These opportunities, however, should not be understood as direct or automatic outcomes, since their realization depends on institutional safeguards, governance mechanisms, transparency, accountability, and human oversight. These opportunities range from automating administrative processes to analyzing public sentiment and enhancing deliberative debate, with the potential to increase civic participation and support more responsive, data-driven policymaking. Each opportunity is examined through a consistent lens: its definition, enabling conditions, and democratic implications.

The taxonomy identifies opportunities and enabling conditions to guide responsible innovation and policy.

\begin{wrapfigure}{r}{0.42\textwidth}
\vspace{-8pt}
\centering
    \includegraphics[width=\linewidth]{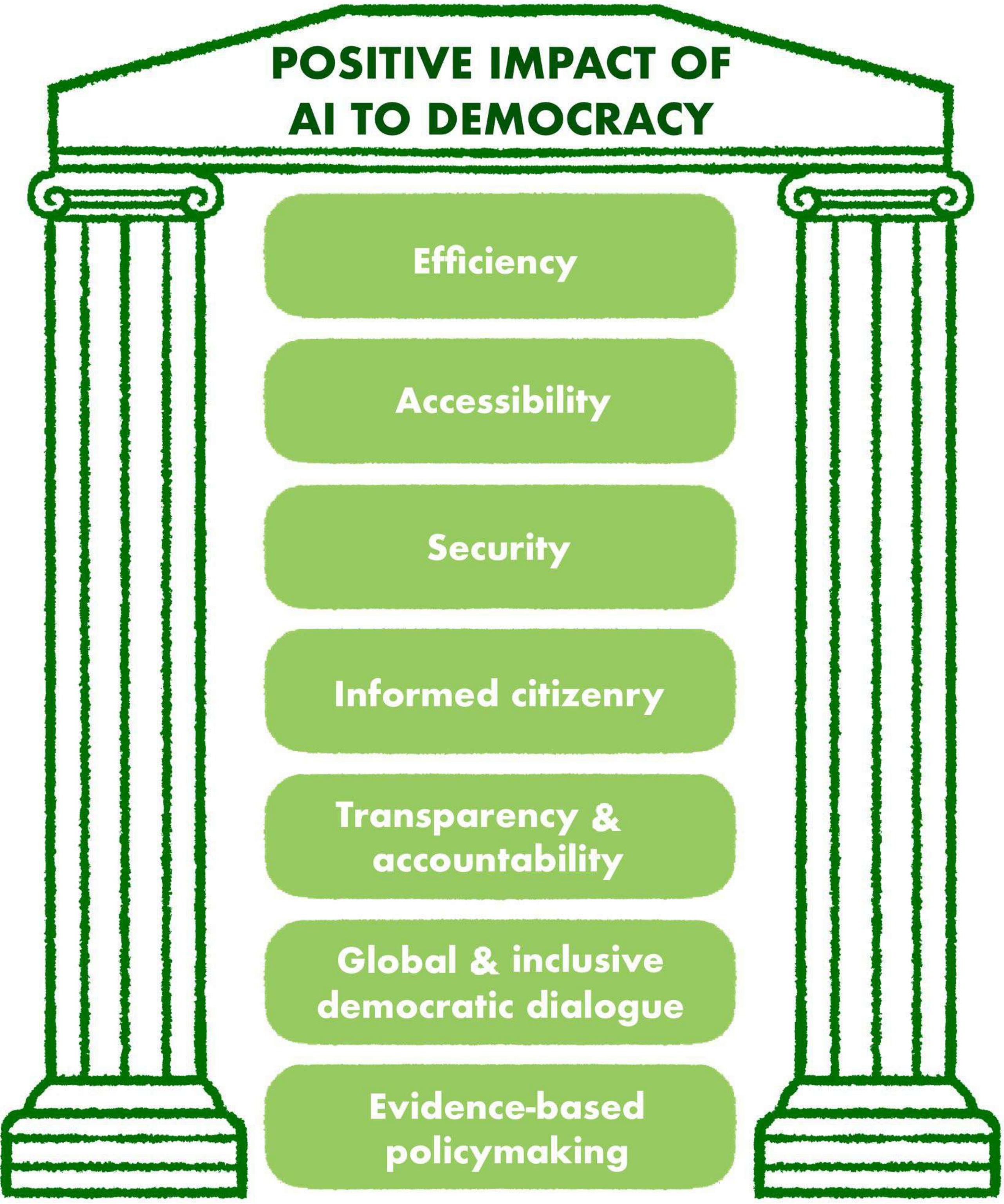}
    \caption{Identified positive contributions of AI to democracy.}
    \label{fig:positiveImpact}
\end{wrapfigure}

This section explores the positive contributions of AI to democracy across key domains. The same intersectional considerations apply to the AIPD taxonomy: positive contributions such as improved access to information, enhanced participation, or better-targeted public services may particularly benefit disadvantaged groups when they are designed and governed with distributive and intersectional concerns explicitly in mind. These areas, derived from the literature review, form the basis of the AIPD taxonomy (AI’s Positive Contributions to Democracy), understood here as a set of conditional democratic opportunities rather than guaranteed outcomes, as illustrated in Fig. \ref{fig:positiveImpact}.


The positive democratic potential of AI also requires more systematic treatment. Recent work on AI-mediated deliberation suggests that AI systems can support participation by summarizing large volumes of citizen input, translating between languages, identifying areas of agreement and disagreement, and helping groups formulate common-ground statements. Tessler et al. \cite{tessler2026mediation}, for example, analyze how AI mediation may contribute to democratic deliberation by addressing tensions between broad participation, meaningful deliberation, and political equality. Similarly, Mariani et al. \cite{mariani2026why} discuss AI-enabled civic tech, e-participation, and parliamentary applications as domains in which AI can strengthen citizen engagement and institutional responsiveness.

These contributions, however, do not follow automatically from technical capability. AI-supported participation strengthens democratic processes only when deliberative procedures remain inclusive, transparent, contestable, and institutionally connected. Rymon \cite{rymon2026representatives} shows that AI may transform representatives from interpreters of citizen preferences into facilitators of expressed preferences and architects of algorithmic decision systems. Whether this transformation enhances democracy depends on clear enabling conditions, including democratic authorization of objectives, transparency at the design stage, accountability mechanisms, reduction of participation barriers, and clear authority frameworks. Accordingly, the AIPD taxonomy should be read not as a catalogue of automatic benefits, but as a set of positive democratic contributions whose realization requires institutional safeguards and governance support.

\Needspace{0.48\textheight}
\noindent\textbf{AIPD1. Efficiency} AI can support democratic administration through the automation of labor-intensive tasks, improving effectiveness, and optimizing resource allocation. This category is limited to improvements in institutional or operational performance.

\noindent \textit{Why it happens:} AI can automate transcription, outreach, scheduling, budgeting, content management, and other routine administrative tasks, while simplifying and personalizing public-service workflows. This can reduce administrative burdens and free staff for citizen-facing work.

\noindent \textit{What it leads to:} These efficiencies may improve resource allocation, service responsiveness, and institutional capacity to address citizens' needs.

\noindent \textit{Literature analysis:} 

Park \cite{park2024theodor} emphasizes AI’s potential to enhance governance, while Soon et al. \cite{soon2024safeguarding} demonstrate how generative tools improve operational efficiency—automating routine tasks and optimizing election logistics to level the playing field for under-resourced candidates. Coeckelbergh \cite{coeckelbergh2024artificial} notes that algorithmic systems support complex decision-making through data processing, and Innerarity \cite{innerarity2024artificial} adds that AI enhances public service efficiency and strategic planning. Finally, Grimmelikhuijsen and Tangi \cite{grimmelikhuijsen2024factors} provide empirical evidence across seven EU countries that these expectations of improved service delivery drive AI adoption among public managers.

\paragraph{\bf{AIPD2. Accessibility}} AI can help make democratic participation more accessible, particularly for individuals and groups with limited financial or technological resources. By reducing barriers to information and participation, AI can enable more citizens to engage with democratic processes. The emphasis is on enabling citizens to reach and use democratic processes that would otherwise remain difficult to access because of practical, linguistic, cognitive, economic, or technological barriers.

\noindent \textit{Why it happens:} AI tools can lower barriers to public information, services, political discussion, and interaction with government platforms, enabling broader democratic participation.

\noindent \textit{What it leads to:} Broader access may diversify representation, reduce power asymmetries, and strengthen institutional inclusiveness, legitimacy, and public trust.

\noindent \textit{Literature analysis:} Existing studies suggest that AI may improve democratic accessibility through two main mechanisms. First, AI can reduce organizational and resource barriers by helping resource-constrained candidates manage campaign logistics, outreach, and strategy more effectively, thereby broadening political representation and facilitating engagement with under-represented groups through tailored messaging and digital channels \cite{soon2024safeguarding}. Second, AI can reduce communicative and informational barriers by supporting strategic decision-making, real-time voter analysis, and multilingual, personalized communication \cite{park2024theodor}. The contribution of AIPD2 is therefore not accessibility in an abstract sense, but the practical reduction of financial, technological, linguistic, and informational obstacles that otherwise restrict democratic participation.

\paragraph{\bf{AIPD3. Security}} 
AI can strengthen the security of democratic systems by supporting threat detection, risk prevention, and the protection of both digital infrastructures and physical environments. While often associated with cybersecurity, AI also plays a growing role in maintaining physical safety when integrated into embodied systems such as autonomous vehicles, drones, and robots. This category is limited to the protection of democratic continuity, integrity, and resilience.

\noindent \textit{Why it happens:} AI can detect threats, anomalies, fraud, cyber manipulation, and physical hazards, supporting faster responses while reducing systemic vulnerabilities and risks arising from human error.

\noindent \textit{What it leads to:} Protecting digital and physical infrastructures may strengthen democratic continuity, institutional resilience, trust, and autonomy, provided that security measures include safeguards against surveillance, coercion, and privacy harms.

\noindent \textit{Literature analysis:} 
Wihbey et al. \cite{wihbey2024ai} highlight AI's role in supporting governance resilience, while Encarnacao et al. \cite{da2023framework} underscore the use of anomaly detection for addressing inequalities such as healthcare or climate disparities. Mannu \cite{sharmaimpact} emphasizes the potential of AI combined with blockchain for authentication, fraud prevention, and secure participation. Vagianos \cite{vagianos2022surveillance} illustrates how real-time surveillance technologies, despite concerns, were used during COVID-19 to improve crisis response. Hupont et al. \cite{hupont2022landscape} highlight how biometric and emotion recognition technologies can introduce safety and security risks, particularly when used in law enforcement, and stress the need for explainable, privacy-preserving systems aligned with the Trustworthy AI framework.

\paragraph{\bf{AIPD4. Informed Citizenry}} An informed citizenry is a cornerstone of democracy, and AI greatly improves access to information using tools such as digital assistants. This category concerns citizens' ability to understand, verify, and contextualize politically relevant information.

\noindent \textit{Why it happens:} Chatbots and AI-powered tools help clarify policies, provide essential information, and facilitate communication with government officials, enabling citizens to stay informed about political developments and engage more effectively in democratic processes. In addition, AI enables personalized content delivery, which can improve access to information relevant to citizens' concerns and support their ability to make informed decisions.

\noindent \textit{What it leads to:} In an era of misinformation, AI can counter false narratives and help ground public debate in facts. This supports trust, political engagement, and citizens' access to accurate and timely information for democratic participation.

\noindent \textit{Literature analysis:} Soon et al. \cite{soon2024safeguarding} note that AI enhances civic engagement by helping citizens stay informed and encouraging participation in democratic processes. Mannu \cite{sharmaimpact} adds that AI-driven personalization allows individuals to navigate complex policy issues through tailored content, while also enabling targeted government communication, even where resources are limited. In addition, AI can contribute to safeguarding public discourse by detecting misinformation, disinformation, and harmful content. Algorithms can flag hate speech and support fact-checking initiatives, helping debates remain evidence-based and fostering a more informed and resilient citizenry.

\paragraph{\bf{AIPD5. Transparency and Accountability}} 
Transparency and accountability are essential to maintain the legitimacy and trustworthiness of democratic institutions. AI can enhance both by monitoring government actions and identifying biases within decision-making processes, supporting more open and accountable decision-making. The emphasis is on making decisions, institutions, or systems open to scrutiny, so that their actions can be examined, justified, challenged, or corrected.

\noindent \textit{Why it happens:} 
AI can analyze large datasets to detect patterns of discrimination in areas such as law enforcement, hiring practices, and access to public services. It also allows governments to track the implementation of policies, providing real-time insight into their effectiveness and areas for improvement. These capabilities can help governments assess whether policies are being applied fairly and efficiently, supporting a higher standard of governance.

\noindent \textit{What it leads to:} By identifying discriminatory patterns and tracking policy performance, AI can support scrutiny, corrective action, equitable services, accountability, and institutional trust.

\noindent \textit{Literature analysis:} 
AI can promote fairness and equity by identifying systemic biases in governance. Encarnacao et al. \cite{da2023framework} highlight how AI can analyze data for discriminatory patterns in areas such as law enforcement and public services, enabling targeted interventions and fairness audits. Matheus et al. \cite{Matheus2021} emphasize AI’s role in tracking policy implementation and public spending, providing real-time transparency and accountability (e.g., AI-powered dashboards that detect irregularities). Jungherr \cite{jungherr2023artificial} adds that algorithmic tools help manage political complexity by measuring public preferences and predicting policy impacts, reinforcing accountability in democratic decision-making. Recent research further shows that transparency and accountability increasingly require technical capacities to monitor synthetic content: generative LLMs and Vision--Language Models (VLMs) can amplify fake news, deepfakes, and manipulated media, but can also support detection, provenance tracking, and content verification \cite{radanliev2025generative, coeckelbergh2025llms, triguero2024general}. This dual role makes technical oversight of synthetic media central to AIPD5.

\paragraph{\bf{AIPD6. Global and Inclusive Democratic Dialogue}} 
AI can strengthen democratic processes globally by fostering cooperation, participation, and communication among democratic societies. The emphasis is on connecting publics, communities, languages, or institutions that would otherwise remain separated within democratic deliberation.

\noindent \textit{Why it happens:} 
AI-powered tools such as translation and coordination systems can bridge linguistic and cultural barriers, but also support cross-border collaboration through shared data infrastructures and digital policy platforms. These tools can facilitate inclusive and transparent discussions among citizens, governments, and international organizations, promoting mutual understanding and coordinated democratic action. Unlike AIPD4, which focuses on citizens’ access to information within national contexts, this dimension addresses the systemic and transnational mechanisms through which AI enables cooperation and coordination among democratic actors at the global level.

\noindent \textit{What it leads to:} 
AI thus may contribute to a more cohesive and participatory global democratic order. By enabling multilingual participation and reducing power asymmetries in communication and access to information, it can strengthen cooperation between democratic nations and support more effective and inclusive governance. This may foster mutual understanding, international stability, and the diffusion of democratic values worldwide.

\noindent \textit{Literature analysis:} This literature frames AI-supported democratic dialogue around three recurrent dimensions: multilingual inclusion, transnational coordination, and normative alignment with democratic values. First, multilingual AI tools can broaden participation by making electoral materials and political communication accessible across linguistic communities \cite{soon2024safeguarding}. Second, secure and personalized digital platforms can support cross-border engagement, policy discussion, and coordination among citizens, governments, and international organizations \cite{sharmaimpact}. Third, the literature emphasizes that these tools strengthen democracy only when they remain aligned with human rights, pluralism, autonomy, and the common good \cite{kan2024artificial, coeckelbergh2024artificial, overton2024overcoming}. Thus, AIPD6 captures AI’s capacity to connect publics and institutions across linguistic, cultural, and territorial boundaries, while remaining conditional on democratic safeguards.

\paragraph{\bf{AIPD7. Evidence-Based Policymaking}} 
AI can support evidence-based policymaking by analyzing data trends, predicting policy outcomes, and assessing public sentiment. The emphasis is on informing which policy options should be adopted, revised, or evaluated, rather than on improving the administrative execution of public tasks.

\noindent \textit{Why it happens:} 
AI can help identify emerging needs, summarize citizen input, and forecast outcomes. Evaluating a policy intervention additionally requires a defined target population, outcome, time horizon, and comparison policy. As L{\'o}pez de Prado and Zoonekynd emphasize~\cite{lopezdeprado2026factormirage}, predictive fit alone cannot justify the selection of adjustment variables: omitted common causes and inappropriate adjustment for common effects can distort causal estimates. Social-media evidence also requires assessment of selection and representation. Where causal effects cannot be identified credibly, results should be presented as descriptive evidence, forecasts, or scenarios, with explicit limits on their policy interpretation.

\noindent \textit{What it leads to:} With reliable, representative data and credible causal assumptions, AI may support more responsive and fair policy while strengthening institutional trust.

\noindent \textit{Literature analysis:} 
AI can support evidence-based policymaking by analyzing data trends, forecasting outcomes, evaluating public sentiment, and processing citizen feedback at scale \cite{duberry2022artificial, sharmaimpact}. These capabilities allow governments to better align policies with societal needs, design more responsive interventions, and anticipate the effects of alternative policy options. AI can also identify patterns in areas such as inequality, supporting targeted public action \cite{da2023framework}.

\section{Risk taxonomy: risks posed by AI to democracy}
\label{sec:risk}

Despite its potential to support democracy, the integration of AI into public life also introduces serious threats to the integrity and sustainability of democratic institutions. If left unaddressed, these risks may erode fundamental principles such as transparency, accountability, and fraternity. To navigate these challenges, a structured taxonomy is essential—not only to raise awareness but also to help policymakers, researchers, and stakeholders identify vulnerabilities and develop targeted mitigation strategies. Accordingly, the taxonomy supports diagnosis, regulation, oversight, public awareness, and preventive action.

This taxonomy engages with the notion of systemic risk articulated in major governance instruments, but reframes it from a political-theoretical perspective centered on democracy. While the DSA \cite{act2022DSA}, the AI Act \cite{act2024AIA}, and the General-Purpose AI (GPAI) Code of Practice \cite{chairs2025} address systemic risks linked to platforms, search engines, and general-purpose AI models, our AIRD taxonomy focuses specifically on how AI can undermine democratic principles, institutional resilience, and democratic governance.

\begin{wrapfigure}{r}{0.42\textwidth}
\vspace{-8pt}
\centering
    \includegraphics[width=\linewidth]{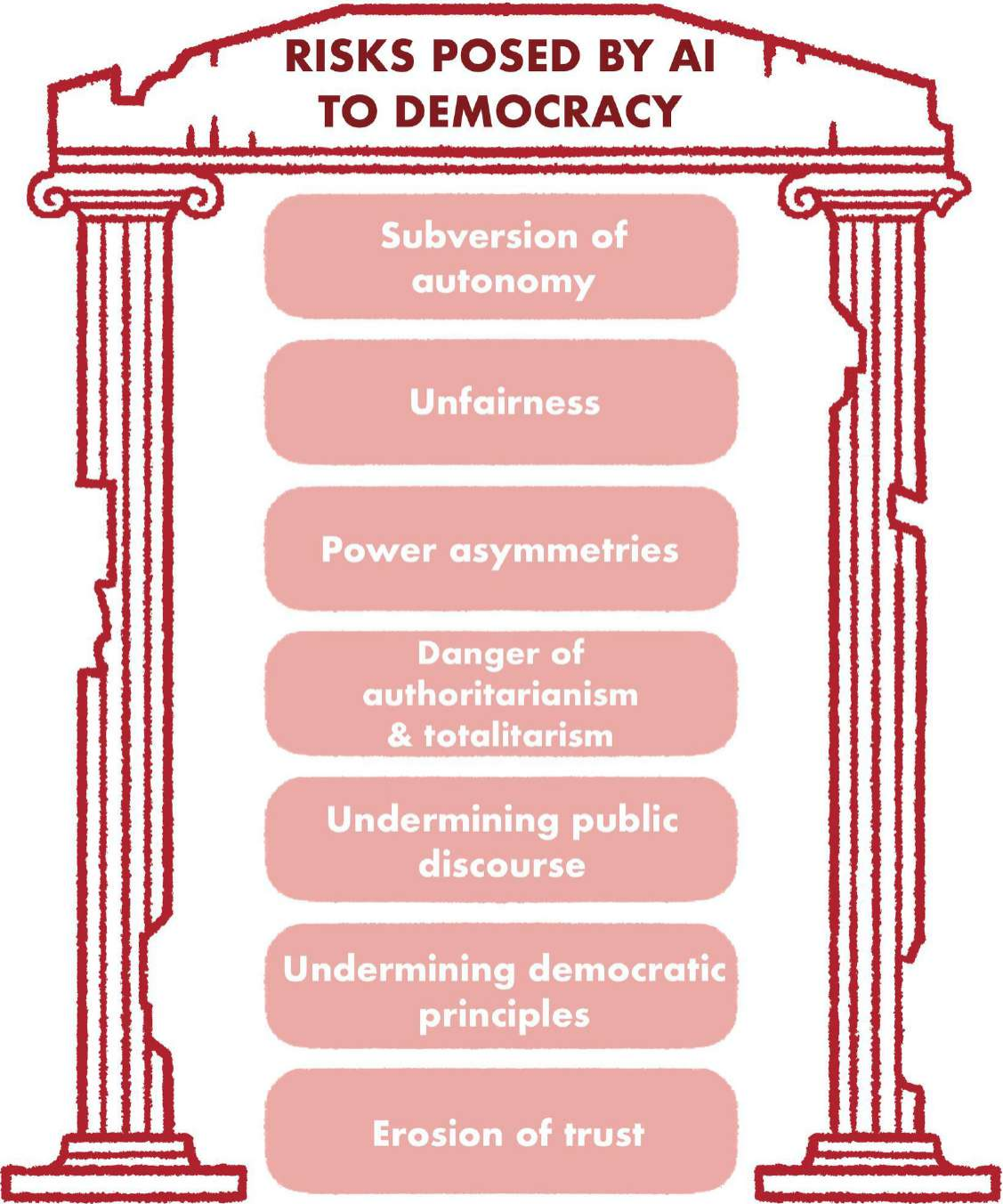}
    \caption{Identified risks posed by AI to democracy.}
    \label{fig:negativeImpact}
\end{wrapfigure}

Building on Coeckelbergh’s work \cite{coeckelbergh2024why} and broader academic discourse, AIRD categorizes risks posed by AI to democracy, as illustrated in Fig.~\ref{fig:negativeImpact}. Although formulated at the level of core democratic principles and institutional risks, many AIRD categories are intersectional: unfairness, exclusion, under-representation, and concentrations of informational or infrastructural power affect citizens differently across gender, race and ethnicity, class, disability, and migration status, often reinforcing pre-existing inequalities. Each risk is analyzed through a structured outline: definition, causes, consequences, and associated literature.

\Needspace{0.48\textheight}
\noindent\textbf{AIRD1. Subversion of Autonomy} Subversion of autonomy refers to the erosion of an individual's ability to make independent decisions free from undue influence or manipulation. The emphasis is on individual agency and decision-making, rather than on the quality of public debate or the level of trust in institutions.

\noindent \textit{Why it happens:} This occurs when AI systems, particularly those optimized for engagement, exploit cognitive biases and manipulate user preferences. Through algorithmic filtering and personalized content, individuals are nudged toward specific viewpoints and behaviors—often without awareness—limiting exposure to diverse information and weakening their capacity for independent thought.

\noindent \textit{What it leads to:} The resulting manipulation diminishes individuals’ ability to make informed choices, undermining democratic participation. As autonomy erodes, so does critical analysis and civic agency, threatening the legitimacy of democratic processes built on free and informed decision-making.

\noindent This risk raises serious concerns for freedom, fraternity, and the rule of law by shaping individual decisions without full awareness or informed consent.\\
\noindent \textit{Literature analysis:} Coeckelbergh \cite{coeckelbergh2024why} links this risk to the erosion of epistemic agency\footnote{Epistemic agency refers to an individual's capacity to control, form, and revise their beliefs in the context of acquiring and processing knowledge \cite{coeckelbergh2025ai}}, as individuals become passive recipients of curated narratives rather than reflective agents. He also describes how “choice architecture” \cite{coeckelbergh2023narrative} can subtly steer behavior while reducing critical awareness, and argues that a lack of hermeneutic responsibility\footnote{Hermeneutic responsibility goes beyond causal explanations to consider broader ethical and social implications \cite{coeckelbergh2020artificial}.} further widens the gap between technical functionality and lived experience, since AI processes data without interpreting human meaning. Nie \cite{nie2024artificial} adds that generative AI amplifies misinformation, further weakening independent judgment; Kan et al. \cite{kan2024artificial} emphasize how surveillance and behavioral tracking, especially through recommendation systems that narrow exposure to diverse views, reduce rational autonomy; and Innerarity \cite{innerarity2024artificial} argues that AI can create an illusion of autonomy by subtly shaping political choices and shifting agency from humans to algorithms, enabling a form of “automated democracy.”

From a causal inference perspective \cite{hernan2020causal, de2023causal}, this dynamic could be modeled through links between optimization objectives, personalization, reduced exposure diversity, and autonomy loss, helping identify interventions that restore autonomy and epistemic agency.

\paragraph{\bf{AIRD2. Unfairness}} 
AI systems can reinforce societal biases, resulting in unfair outcomes that perpetuate inequality. The emphasis is on unequal treatment, discriminatory outcomes, and the distribution of burdens or benefits across social groups.

\noindent \textit{Why it happens:} 
This risk stems from biased training data, opaque models, and inadequate oversight in sensitive domains such as hiring, policing, and public services. Human prejudices embedded in datasets, combined with homogeneous development teams and insufficient impact assessments, allow biased systems to operate unchecked.

\noindent \textit{What it leads to:} 
Discriminatory outputs disproportionately harm marginalized groups, undermining fairness, representation, and non-discrimination, key pillars of democratic equality.

\noindent This risk directly affects equality and fraternity by reinforcing discrimination and deepening structural injustice.

\noindent \textit{Literature analysis:} Research on algorithmic unfairness identifies biased, incomplete, historical, or proxy data as key sources of democratic harm, since they can reproduce existing inequalities in domains such as hiring, lending, policing, public services, education, and judicial decision-making \cite{da2023framework, winter2022challenges, grossi2024enabling, duberry2022artificial}. Personalization, profiling, and opaque automated decision-making can further make unequal treatment difficult to detect, contest, or correct, especially for groups differentiated by race, education, socioeconomic status, or political behavior \cite{overton2024overcoming, kucirkova2023beyond}. Thus, unfairness is not only a technical problem of biased prediction, but also a democratic problem concerning the unequal distribution of burdens, opportunities, and political agency across social groups.

López de Prado \cite{de2023causal} suggests that causal modeling helps specify empirical models correctly and avoid biases from non-causal associations; applied here, it could test whether discrimination primarily stems from biased training data or flawed modeling choices, informing whether fairness audits should target data practices, algorithmic adjustments, or both.

\paragraph{\bf{AIRD3. Power Asymmetries}} 
AI exacerbates existing power imbalances by concentrating control over data, infrastructure, and decision-making in the hands of a few corporations and governments. The emphasis is on structural concentration of influence and control, rather than on the direct use of AI for coercion, censorship, or authoritarian governance.

\noindent \textit{Why it happens:} 
Monopolization of AI resources—data, computing power, and governance—reduces public influence and weakens democratic accountability. Opaque AI governance and limited civic participation further entrench corporate and state dominance.

\noindent \textit{What it leads to:} 
These asymmetries hinder public oversight, restrict innovation, and embed systemic bias, ultimately threatening democratic governance and social equity.

\noindent This risk affects fairness, equality, the rule of law, and tolerance by centralizing power among those who control AI systems.

\noindent \textit{Literature analysis:}
AI enables actors with access to large-scale data and infrastructure to manipulate public discourse, surveil populations, and control information flows. Calzada \cite{Calzada2024democratic} describes the rise of “data-opolies,” where data extractivism and content curation concentrate political influence. Coeckelbergh \cite{coeckelbergh2024artificial} links this to a “democracy deficit” in AI governance, while Buhmann et al. \cite{buhmann2023deep} show how strategic, expert, and emergent opacity consolidate technocratic control and reduce citizens’ ability to question AI-shaped decisions.

López de Prado \cite{de2023causal} argues that causal AI can help attribute effects to causes; applied here, such methods could trace how data concentration reduces civic power through asymmetric policy influence, institutional opacity, and weakened democratic accountability.

\paragraph{\bf{AIRD4. Danger of Authoritarianism and Totalitarianism}} 
AI can be exploited to reinforce authoritarian regimes by enabling enhanced surveillance, censorship, and control over public discourse. The emphasis is on coercive or repressive uses of AI that restrict political freedoms, rather than on broader power concentration or gradual democratic erosion within formally democratic systems.

\noindent \textit{Why it happens:} 
Tools such as facial recognition, predictive policing, automated censorship, and AI-driven propaganda allow governments to suppress dissent, manipulate narratives, and monitor citizens at scale. The integration of AI in governance, often lacking oversight, provides regimes with unprecedented power to enforce compliance and stifle opposition.

\noindent \textit{What it leads to:} 
These technologies undermine democratic principles by restricting freedoms, eroding transparency, and consolidating state control. As power becomes increasingly centralized, regime resilience grows while avenues for political pluralism diminish.

\noindent This risk threatens fairness, equality, the rule of law, and tolerance by enabling coercive governance and silencing civic agency.

\noindent \textit{Literature analysis:}
Labuz et al. \cite{labuz2024way} and Vagianos et al. \cite{vagianos2022surveillance} show how AI enables authoritarian regimes to monitor, censor, and control populations, with China’s “Great Firewall” and social credit system illustrating AI-enabled tracking and enforced conformity. Labuz et al. also warn that AI-driven disinformation through deepfakes\footnote{AI-generated media that falsely appear authentic.} and bots\footnote{Scripts that automate online behaviors.} can distort elections and manipulate opinion. Encarnacao et al. \cite{da2023framework}, Innerarity \cite{innerarity2024artificial}, and Nie \cite{nie2024artificial} further link AI opacity and “algocracy” to reduced accountability, weaker public oversight, and unchecked state power.


Recent work on technofascism further extends this concern by linking AI-enabled authoritarianism not only to state surveillance and coercion, but also to the concentration of technological power in Big Tech infrastructures. Coeckelbergh \cite{coeckelbergh2026technofascism} argues that AI, platform monopolization, behavioral nudging, and infrastructural control may support new authoritarian tendencies even beyond traditionally authoritarian regimes. This reinforces the distinction between AIRD3 and AIRD4: while AIRD3 captures structural power asymmetries, AIRD4 captures the use of AI-enabled power for coercive, repressive, or authoritarian forms of political control.

\paragraph{\bf{AIRD5. Undermining Public Discourse}} 
AI-driven platforms, particularly recommender systems, increasingly shape public discourse by promoting emotionally charged and polarizing content over balanced, fact-based information. This not only fragments the public sphere but also erodes institutional trust by amplifying narratives that delegitimize democratic institutions and fuel public cynicism. The emphasis is on degrading the public information environment and deliberative debate, rather than on the structural weakening of democratic principles or the downstream loss of trust.

\noindent \textit{Why it happens:} 
Optimized for engagement, AI algorithms amplify content likely to trigger emotional reactions, reinforcing confirmation bias and creating echo chambers. This, combined with the spread of misinformation and synthetic media, distorts the information ecosystem and undermines shared understanding.

\noindent \textit{What it leads to:} 
Public deliberation deteriorates, polarization deepens, and consensus becomes harder to reach. Citizens' capacity for informed decision-making is weakened, directly threatening the democratic principle of tolerance.

\noindent \textit{Literature analysis:} 
The literature on AI and public discourse highlights three mechanisms of democratic disruption. First, engagement-oriented recommender systems can amplify polarizing or emotionally charged content, discouraging critical dialogue and weakening exposure to diverse viewpoints \cite{sheikh2024data, kan2024artificial, coeckelbergh2024why}. Second, personalization can fragment the public sphere by creating ideologically segmented information environments in which citizens receive increasingly different versions of political reality. Third, data-driven manipulation, including micro-targeting\footnote{The 2018 Cambridge Analytica scandal and political campaigns such as Brexit and the 2020 U.S. election can be examples of this \cite{sheikh2024data}.}, bots, disinformation, and synthetic media, can weaponize personal data to shape opinion and distort deliberation \cite{jones2025don, duberry2022artificial}. These mechanisms show that AIRD5 is not limited to isolated false content, but concerns the broader degradation of the information environment in which democratic debate takes place. Recent work on AI in campaigns, elections, movements, and deliberation similarly shows that AI-mediated democratic effects cut across electoral communication, political mobilization, and public reasoning, reinforcing the need to analyze public discourse as a broader democratic environment rather than as a narrow problem of isolated false content \cite{flanigan2026artificial}.


Recent developments in generative and agentic AI reinforce the urgency of the AIRD taxonomy. Schroeder et al. \cite{schroeder2026malicious} describe malicious AI swarms as coordinated networks of AI-controlled agents capable of maintaining persistent identities, adapting to engagement signals, coordinating across platforms, and fabricating apparent consensus. These systems intensify existing risks of misinformation, manipulation, erosion of public trust, and polarization because they can operate at greater scale, speed, and contextual sophistication than earlier botnets or manually coordinated influence operations. Within the AIRD framework, such swarms are not confined to a single category: they may primarily instantiate risks to informed autonomy and trust, while also generating secondary risks related to power asymmetries, authoritarian manipulation, and the degradation of democratic deliberation.
The electoral context illustrates the governance relevance of this risk. Kalsnes and Schia \cite{kalsnes2026protecting} show that several countries introduced codes of conduct or similar guidelines for generative AI use during elections, but that these instruments often remain limited by weak enforceability, inconsistent definitions of harmful use, and insufficiently clear accountability mechanisms. This supports the need for a taxonomy capable of distinguishing different democratic harms and linking them to responsible actors and governance responses. In electoral settings, AIRD categories can help clarify whether a given AI use primarily threatens information integrity, voter autonomy, equal political competition, institutional trust, or accountability.

\paragraph{\bf{AIRD6. Undermining Democratic Principles}} 
AI can undermine democratic principles when it weakens the normative and institutional conditions that sustain democratic self-government, including equality, participation, accountability, the rule of law, representation, and democratic legitimacy. The emphasis is therefore not on a specific information disorder or trust failure, but on the broader erosion of the principles that make collective democratic decision-making possible.

\noindent \textit{Why it happens:} 
AI can undermine democratic principles when democratic functions are delegated to opaque, weakly accountable, or insufficiently contestable systems. This occurs when AI shapes participation, public administration, representation, or institutional decision-making without adequate transparency, human oversight, legal safeguards, or channels for public contestation.

\noindent \textit{What it leads to:} 
The result is a loss of democratic legitimacy: citizens have fewer meaningful opportunities to understand, contest, or influence decisions, while institutions become harder to justify democratically.

\noindent This risk differs from AIRD4 in that it captures internal democratic erosion rather than external authoritarian control.

\noindent \textit{Literature analysis:} 
Innerarity \cite{innerarity2024artificial} argues that AI can foster ideological fragmentation, public confusion, and manipulation by privileging engagement over truth in digital spaces. Jungherr \cite{jungherr2023artificial} notes that AI may improve governance, but also shift power from voters to technocratic actors and undermine electoral integrity through content moderation and micro-targeting. This concern is reinforced by recent work showing that AI may transform the role of democratic representatives by reshaping how political judgment, delegation, and representation are mediated algorithmically \cite{rymon2026representatives}.
Park \cite{park2024theodor} adds that AI threatens both negative and positive freedom through surveillance and biased algorithms, restricting autonomy and participation. Beckman et al. \cite{beckman2024artificial} warn that opaque AI decision-making in public administration undermines institutional legitimacy unless embedded within transparent and accountable frameworks. Without such safeguards, AI governance may bypass democratic deliberation, reduce contestability, and weaken public accountability.

\paragraph{\bf{AIRD7. Erosion of Trust}} 
Trust grounds democratic legitimacy, information credibility, and social cohesion. AI technologies can erode this trust when they make it harder for citizens to distinguish reliable from unreliable information, to identify accountable institutional actors, or to rely on their own judgment in AI-mediated environments. The emphasis is on the loss of confidence in information, institutions, electoral processes, and personal judgment, rather than on the initial distortion of public discourse or the broader erosion of democratic principles.

\noindent \textit{Why it happens:} 
Erosion of trust can be understood in two related dimensions. First, AI-generated or AI-amplified content can weaken epistemic trust by making citizens less confident about what they see, hear, or read. This effect is intensified by generative AI, deepfakes, synthetic media, bots, and micro-targeted communication, which can blur the boundary between authentic and manipulated information. Second, institutional trust is weakened when AI systems operate opaquely, when responsibility for their outputs is unclear, or when citizens perceive that public decisions, electoral communication, or administrative processes are being shaped by systems they cannot understand or contest.

\noindent \textit{What it leads to:} 
This twofold erosion of trust affects confidence in media, electoral systems, public institutions, and one's own capacity for political judgment. As trust declines, citizens may become more vulnerable to cynicism, disengagement, polarization, or conspiratorial interpretations of public life. In this way, AI can distort democratic dynamics, weaken social cohesion, and reduce the credibility of the deliberative and institutional processes on which democratic legitimacy depends. 

\noindent \textit{Literature analysis:} 
Coeckelbergh \cite{coeckelbergh2023democracy, coeckelbergh2025llms} argues that AI-driven echo chambers and LLMs erode epistemic agency, amplify misinformation, and weaken factual discourse. Denemark \cite{denemark2024risk} adds that AI can distort perceived reality and undermine public trust, while Bareis \cite{bareis2024trustification} shows how AI chatbots can emotionally manipulate users and degrade trust in institutions and in one's own judgment. Taken together, these contributions show that AIRD7 is not merely a problem of false information, but a broader democratic risk: when citizens cannot reliably identify trustworthy information, responsible actors, or contestable procedures, both institutional legitimacy and democratic agency are weakened.

\section{Trustworthy AI requirements for AI risk mitigation}
\label{sec:maping}
To protect democratic values in the age of AI, conceptual insights must be paired with actionable governance tools. Building on Section II.C and on the AIRD risks described in Section IV, this section maps each AIRD category to the European Commission's Trustworthy AI requirements \cite{hleg2019ethics}. These requirements translate ethical principles into design standards and oversight mechanisms for AI systems.

Trustworthy AI provides a practical bridge between high-level values and concrete governance practices. Díaz-Rodriguez et al. \cite{diaz2023connecting} describe it as a connective architecture linking transparency, explainability, accountability, and human autonomy to responsible AI systems and regulatory strategies, while Li et al. \cite{li2023trustworthy} emphasize its operationalization across the AI lifecycle. Integrating these requirements into the AIRD taxonomy strengthens it as a tool not only for identifying democratic risks, but also for defining mitigation measures and supporting democratic resilience.

\begin{figure*}[htbp!]
    \centering
    \includegraphics[width=1\textwidth]{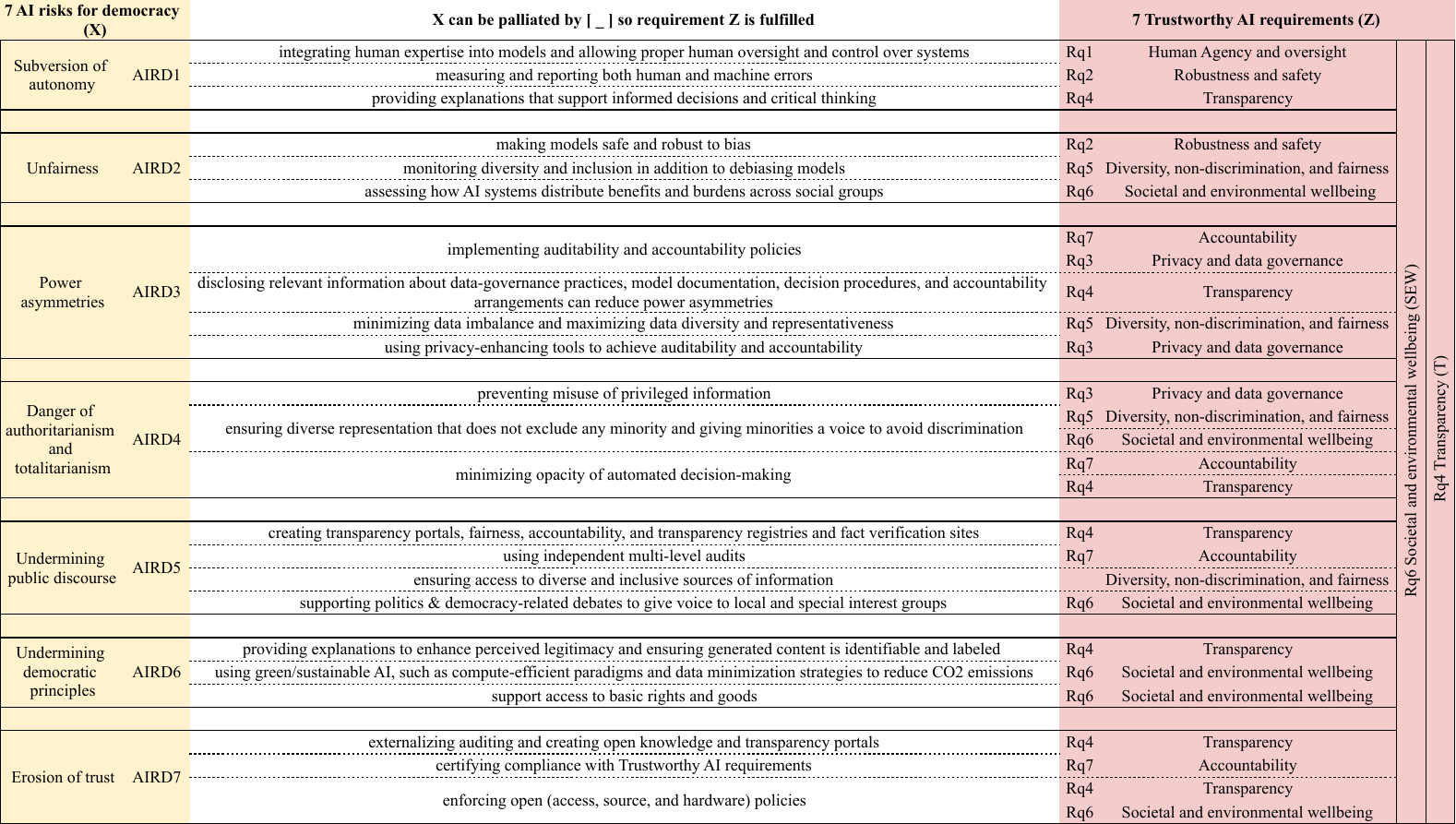}
 \caption{Mapping AI risks to democracy onto Trustworthy AI requirements and associated mitigation measures.}
    \label{fig:2ndMapping}
\end{figure*}

The mapping below is conceptual rather than empirically validated, but it is informed by the literature and by widely accepted links between ethical AI governance and democratic resilience. The mapping and corresponding recommendations are summarized in Fig.~\ref{fig:2ndMapping} and detailed below.

\begin{enumerate}
    \item \textit{Subversion of autonomy}:
    \begin{itemize}
        \item \textit{Subversion of autonomy} can be mitigated by integrating human expertise into models and allowing proper human supervision and control mechanisms over systems, so the requirement for \textit{human agency and oversight} is met. This ensures that human decision-making remains central and prevents the complete delegation of control to machines, thus safeguarding individual autonomy.
        \item \textit{Subversion of autonomy} can be reduced by measuring and reporting both human and machine errors so \textit{robustness and safety} requirements are met. By implementing mechanisms to maintain performance and addressing both human and machine errors, the system becomes more reliable, thereby reinforcing safety and autonomy.
        \item \textit{Subversion of autonomy} can be minimized by providing explanations that support informed decisions and critical thinking so the requirement for \textit{transparency} is met. Providing explanations of AI outputs fosters transparency, enabling individuals to make informed choices while preserving their autonomy.
    \end{itemize}
    
    \item \textit{Unfairness}: 
    \begin{itemize}
        \item \textit{Unfairness} can be mitigated by monitoring diversity and inclusion in addition to debiasing models so \textit{diversity, non-discrimination, and fairness} are realized. This emphasizes the importance of continuous monitoring to reduce biases that could otherwise lead to discriminatory outcomes.
        \item \textit{Unfairness} can be mitigated by making models safe and robust to bias and adversarial attacks so \textit{robustness and safety} are met. Ensuring that models are robust to bias and adversarial attacks enhances fairness by minimizing vulnerabilities that can arise from the risk of unfair treatment. 
        \item \textit{Unfairness} can also be mitigated by assessing how AI systems distribute benefits and burdens across social groups, so \textit{societal and environmental wellbeing} is addressed. This requirement is relevant because discriminatory or exclusionary outcomes can undermine social cohesion, equal access to democratic participation, and the broader societal conditions that support democratic legitimacy.
    \end{itemize}

    \item \textit{Power asymmetries}:
    \begin{itemize}
        \item \textit{Power asymmetries} can be minimized by implementing accountability and auditability policies so \textit{accountability} is met. Auditability and accountability mechanisms ensure that those in power can be held responsible for their actions, preventing the abuse of technological power.
        \item \textit{Power asymmetries} can be mitigated by disclosing relevant information about data-governance practices, model documentation, decision procedures, and accountability arrangements so \textit{transparency} is achieved. Open transparency portals and accessible explanations of model-assisted decisions reduce the risk of power imbalances, as individuals and organizations can understand how decisions are made.
        \item \textit{Power asymmetries} can be mitigated by minimizing asymmetries arising from AI systems through the implementation of \textit{diversity, non-discrimination, and fairness} policies. Minimizing AI-driven automated decisions can help ensure no group holds disproportionate power, fostering fairness and equality.
        \item \textit{Power asymmetries} can be mitigated by using privacy-enhancing tools to achieve auditability and accountability so \textit{privacy and data governance} are attained. Privacy-enhancing tools protect personal data, ensuring that systems maintain data integrity, so the misuse of power is prevented.
    \end{itemize}

    \item \textit{Danger of authoritarianism and totalitarianism}:
    \begin{itemize}
        \item \textit{Danger of authoritarianism and totalitarianism} can be mitigated by preventing the misuse of privileged information so the requirement for \textit{privacy and data governance} is met. This prevents the misuse of sensitive information, which could otherwise concentrate power and control in the hands of a few.
        \item \textit{Danger of authoritarianism and totalitarianism} can be mitigated by ensuring diverse representation (as in power asymmetries) and giving minorities a voice so \textit{diversity, non-discrimination, and fairness} and \textit{societal and environmental wellbeing} are met. Ensuring diverse representation and giving minorities a voice counters the risk of authoritarian control, supporting fairness and social wellbeing.
        \item \textit{Danger of authoritarianism and totalitarianism} can be mitigated by minimizing opacity of automated decision-making so \textit{transparency and accountability} are achieved. Transparency in algorithmic decision-making reduces the risks of authoritarianism, ensuring that citizens are provided with insights into automated processes that affect them.
    \end{itemize}

    \item \textit{Undermining public discourse}:
    \begin{itemize}
        \item \textit{Undermining public discourse} can be mitigated by using independent, multi-level auditing processes so \textit{accountability} is met. External audits ensure that the systems responsible for informing and shaping public discourse remain objective and accountable, reducing the potential for manipulation.
        \item \textit{Undermining public discourse} can be prevented by creating transparency portals and records so \textit{transparency} is met. Transparency portals and registries enable the public to track algorithmic decisions, preventing the erosion of public discourse through hidden manipulations.
        \item \textit{Undermining public discourse} can be mitigated by promoting access to diverse, representative sources of information and inclusive content curation, so \textit{diversity, non-discrimination, and fairness} are met. Ensuring exposure to pluralistic viewpoints helps reduce echo chambers and polarization, supporting a more balanced and equitable democratic debate.
        \item \textit{Undermining public discourse} can be minimized by supporting politics and democracy-related fact verification sites so \textit{societal and environmental wellbeing} are attained. Fact-checking tools and political transparency ensure that misinformation does not undermine the integrity of democratic discourse.
    \end{itemize}

    \item \textit{Undermining democratic principles}:
    \begin{itemize}
        \item \textit{Undermining democratic principles} can be mitigated by providing explanations to enhance perceived legitimacy and ensuring content is identifiable and labeled so \textit{transparency} is met. Providing explanations improves the perceived legitimacy of decisions made by AI systems, fostering transparency in democratic processes.
        \item \textit{Undermining democratic principles} can be reduced by using green/sustainable AI and data minimization strategies so \textit{societal and environmental wellbeing} is realized. Green AI and sustainable AI practices align technological development with democratic values, promoting long-term societal and environmental wellbeing.
        \item \textit{Undermining democratic principles} can be mitigated by widening access to fundamental rights and goods so \textit{societal and environmental wellbeing} is met. Listening to every stakeholder and ensuring that AI systems promote access to basic resources helps ensure that no group is excluded, reinforcing democratic principles. 
    \end{itemize}

    \item \textit{Erosion of trust}:
    \begin{itemize}
        \item \textit{Erosion of trust}  can be mitigated by commissioning independent external audits and by creating open knowledge and transparency portals, thereby supporting transparency. 
        \item \textit{Erosion of trust} can be minimized by certifying compliance with Trustworthy AI requirements so \textit{accountability} is met. Certification ensures that AI systems meet established standards, enhancing public trust by confirming accountability and ethical compliance.
        \item \textit{Erosion of trust} can be mitigated by enforcing open policies (access, source, and hardware) so \textit{transparency} and \textit{societal and environmental wellbeing} are achieved. Open policies make AI systems more accessible and understandable, while supporting broader societal conditions for public trust, inclusion, and democratic accountability. Open policies ensure that the AI systems' inner workings are transparent, mitigating trust erosion by providing public access to the tools and data used.
    \end{itemize}
\end{enumerate}

As can be seen in our analysis mapping requirements for trustworthy AI to AI risks to be mitigated (Fig.  \ref{fig:2ndMapping}), \textit{Societal and Environmental Wellbeing} (\textit{Rq6}) and \textit{Transparency} (\textit{Rq4}) recur across several AIRD categories, although this frequency should not be interpreted as a normative ranking of the Trustworthy AI requirements.

\section{Reflections, Contextualization, and Synthesis}
\label{sec:reflect}

This section examines the framework's operational use, contextual adaptation, democratic trade-offs, lessons, limitations, and links to Trustworthy AI requirements.

\subsection{Operational Use of the Framework}
\label{subsec:operational_use}

The AIRD/AIPD framework operates in four steps: (1) describe the phenomenon's democratic context, affected actors, and institutional setting; (2) assign primary and, where effects overlap, secondary AIRD and/or AIPD categories; (3) map them to relevant Trustworthy AI requirements; and (4) define mitigation or promotion measures. It thereby connects democratic effects to governance responses without serving as a compliance checklist, consistent with mappings of the AI--democracy field in terms of risks, opportunities, actors, and interventions \cite{george2026mapping}.

Assessments should record the proposed mechanism, evidence, uncertainty, responsible actor, measurable democratic outcome, feasible comparison, and review point. Electoral-misinformation evaluation should test whether provenance labels improve factual understanding while monitoring mislabeling and effects on legitimate expression; public-service evaluation should compare processing time, erroneous denials, appeal outcomes, and distributional effects with the existing process. These criteria guide evaluation but do not establish intervention effectiveness.

Table~\ref{tab:operational_mapping} links the seven AIRD categories to diagnostic criteria, Trustworthy AI requirements, governance responses, and responsible actors. Tables~\ref{tab:worked_scenario} and~\ref{tab:worked_scenario_public_services} apply this logic to electoral misinformation and AI-assisted public-service allocation.

The AIRD and AIPD categories should therefore not be interpreted as a binary distinction between harmful and beneficial AI applications. Rather, the framework is intended as an analytical and governance tool for identifying democratic trade-offs, anticipating possible tensions between risks and contributions, and designing interventions that maximize democratic gains while mitigating democratic harms. Governance responses should be selected proportionately, considering the severity, likelihood, scale, and reversibility of democratic harms, as well as institutional capacity and implementation costs.

To avoid overstating the symmetry of the operational framework, we clarify its scope as follows. The paper provides a dual taxonomy of both AIRD and AIPD, but the operational layer is intentionally developed in different ways for each side. AIRD is operationalized primarily as a risk-mitigation framework: it identifies democratic harms, links them to Trustworthy AI requirements, and proposes corresponding governance responses and responsible actors. AIPD is operationalized as an enabling-conditions framework: it identifies ways in which AI may support democratic values, while specifying the institutional, technical, and normative conditions under which such contributions can be realized responsibly. Thus, the framework is dual in its taxonomic structure and differentiated in its operational emphasis.

\begin{table}[!ht]
\centering
\caption{Compact operational mapping of AIRD categories}
\label{tab:operational_mapping}
\footnotesize
\setlength{\tabcolsep}{3pt}
\renewcommand{\arraystretch}{1.15}
\begin{tabularx}{\textwidth}{|Y|Y|Y|Y|Y|}
\hline
\textbf{Category} & \textbf{Diagnostic focus} & \textbf{Trustworthy AI requirements} & \textbf{Governance response} & \textbf{Responsible actors} \\
\hline
\textbf{AIRD1. Subversion of Autonomy} & Manipulation of citizens' choices & Human agency; technical robustness; transparency & Disclosure, contestability, and anti-manipulation safeguards & Regulators; platforms; providers; auditors \\
\hline
\textbf{AIRD2. Unfairness} & Biased outcomes or exclusion from democratic participation & Technical robustness; fairness; societal wellbeing & Bias assessment, representative data, accessibility, and appeal procedures & AI providers; public agencies; auditors; civil society \\
\hline
\textbf{AIRD3. Power Asymmetries} & Concentrated informational, economic, or institutional power & Data governance; transparency; fairness; accountability & Audit trails, transparency portals, and independent oversight & Regulators; platforms; AI providers; auditors; civil society \\
\hline
\textbf{AIRD4. Danger of Authoritarianism and Totalitarianism} & Use of AI for surveillance, coercion, or centralized political control & Data governance; transparency; fairness; societal wellbeing; accountability & Legal limits on surveillance, human rights impact assessments, independent supervision, and redress mechanisms & Regulators; courts; public institutions; oversight bodies; civil society \\
\hline
\textbf{AIRD5. Undermining Public Discourse} & Misinformation, polarization, or degradation of democratic deliberation & Transparency; fairness; societal wellbeing; accountability & Provenance, platform audits, fact-checking, and rapid reporting & Platforms; regulators; civil society; electoral authorities \\
\hline
\textbf{AIRD6. Undermining Democratic Principles} & Weakened participation, pluralism, representation, or accountability & Transparency; societal wellbeing & Participatory governance, democratic impact assessments, public consultation, and institutional transparency duties & Public institutions; regulators; electoral authorities; civil society \\
\hline
\textbf{AIRD7. Erosion of Trust} & Opacity or declining confidence in democratic institutions and information environments & Transparency; societal wellbeing; accountability & Certification, oversight, documentation, reporting, and public communication & Regulators; public institutions; certification bodies; auditors \\
\hline
\end{tabularx}
\end{table}

\begin{table}[!ht]
\centering
\caption{Worked Example: AI-Generated Electoral Misinformation}
\begin{tabularx}{\columnwidth}{|Y|Y|}
\hline
\textbf{Step} & \textbf{Application} \\
\hline
\textbf{Observed Issue} & Synthetic media falsely attribute statements to a candidate before an election. \\
\hline
\textbf{AIRD Classification} & AIRD5; AIRD7; and, with targeted profiling, AIRD1. \\
\hline
\textbf{Trustworthy AI Requirements} & Rq4 \textit{Transparency}; Rq7 \textit{Accountability}; and Rq3 \textit{Privacy and Data Governance}. \\
\hline
\textbf{Governance Responses} & Provenance labels, ad repositories, rapid audits, reporting channels, and electoral authority--platform cooperation. \\
\hline
\end{tabularx}
\label{tab:worked_scenario}
\end{table}

\begin{table}[!ht]
\centering
\caption{Worked Example: AI-Assisted Public Service Allocation}
\begin{tabularx}{\columnwidth}{|Y|Y|}
\hline
\textbf{Step} & \textbf{Application} \\
\hline
\textbf{Observed Issue} & AI-assisted allocation may improve efficiency but create bias, opacity, weak contestability, and unequal group effects. \\
\hline
\textbf{AIPD Classification} & AIPD1 \textit{Efficiency}; AIPD7 \textit{Evidence-Based Policymaking}; and, where AI reduces access barriers, AIPD2 \textit{Accessibility}. \\
\hline
\textbf{Associated AIRD Risks} & AIRD2 \textit{Unfairness}; AIRD3 \textit{Power Asymmetries}. \\
\hline
\textbf{Trustworthy AI Requirements} & Rq1; Rq2; Rq3; and Rq4. \\
\hline
\textbf{Governance Responses} & Human oversight, bias and robustness tests, data safeguards, explanations, appeals, outcome monitoring, and independent audits. \\
\hline
\end{tabularx}
\label{tab:worked_scenario_public_services}
\end{table}

Together, these examples show that the framework supports both risk-oriented and mixed assessments, provided that overlapping categories are made explicit and linked to appropriate safeguards and accountability mechanisms.

\subsection{Global Applicability and Contextualization}

Although grounded in the EU Trustworthy AI framework and liberal-democratic values, AIRD/AIPD is not a rigid template: its local calibration must reflect political, legal, and institutional differences, including constitutional design, conceptions of collective autonomy and individual rights, regulatory capacity, and auditing infrastructure. This is especially important in Global South democracies where technological dependency, limited data sovereignty, or under-resourced oversight may constrain AI governance. Such adaptability supports comparative research, national AI strategies, and context-sensitive regulation.

Table~\ref{tab:rq-frequency} reports recurrence across the seven AIRD categories: \textit{Transparency (Rq4)} appears in six, \textit{Societal and Environmental Wellbeing (Rq6)} in five, and both \textit{Accountability (Rq7)} and \textit{Diversity, Non-Discrimination, and Fairness (Rq5)} in four. Rq6 is not a secondary concern: it links democratic resilience to the social and ecological conditions sustaining institutions, trust, and legitimacy. Sťahel \cite{stahel2026ai} reinforces its relevance by identifying democratic tensions associated with insufficient attention to AI's infrastructural, energetic, and environmental costs.


These counts describe the authors' mapping, not the prevalence or severity of harms, the requirements' normative importance, or intervention effectiveness. Prioritization should instead reflect context, affected groups, evidence, feasibility, and the consequences of action or inaction; future research should test whether governance measures change democratic outcomes.

\begin{table}[!t]
\centering
\caption{Frequency of Trustworthy AI requirements across AIRD risks.}
\label{tab:rq-frequency}
\footnotesize
\setlength{\tabcolsep}{3pt}
\renewcommand{\arraystretch}{0.9}
\begin{tabular}{p{0.13\columnwidth}p{0.58\columnwidth}c}
\hline
\textbf{Req.} & \textbf{Mapped AIRD risks} & \textbf{Freq.} \\
\hline
Rq4 & AIRD1, AIRD3, AIRD4, AIRD5, AIRD6, AIRD7 & 6/7 \\
Rq6 & AIRD2, AIRD4, AIRD5, AIRD6, AIRD7 & 5/7 \\
Rq7 & AIRD3, AIRD4, AIRD5, AIRD7 & 4/7 \\
Rq5 & AIRD2, AIRD3, AIRD4, AIRD5 & 4/7 \\
Rq3 & AIRD3, AIRD4 & 2/7 \\
Rq2 & AIRD1, AIRD2 & 2/7 \\
Rq1 & AIRD1 & 1/7 \\
\hline
\end{tabular}
\end{table}

\subsection{Applying the Framework to LLMs and Frontier AI}

As LLMs and other general-purpose AI (GPAI) systems increasingly mediate civic information flows, they provide a relevant test case for the AIRD/AIPD framework. Their high capability, opacity, and scale can amplify democratic risks and contributions already captured by the proposed taxonomies, while raising specific governance concerns for frontier AI \cite{radanliev2025frontier}. Although included under the GPAI category in the AI Act, LLMs merit distinct attention due to their rapid social diffusion and their role in shaping information access, public debate, and civic interaction \cite{triguero2024general}.

This application of the AIRD/AIPD framework to LLMs does not seek to explain their internal functioning. Rather, given the opacity of LLMs and other deep learning systems, it focuses on observable democratic effects: generated outputs, deployment contexts, governance arrangements, and impacts on information quality, participation, accountability, and trust. While opacity does not prevent effects-based democratic assessment, it may limit causal attribution and the diagnosis of the internal mechanisms responsible for those effects. In this way, the framework helps make LLM-related democratic impacts explicit, comparable, and actionable by linking observable effects to AIRD/AIPD categories, Trustworthy AI requirements, and suitable governance responses.

First, LLMs can undermine autonomy and trust (AIRD1, AIRD7) by generating persuasive but unverifiable content and, when combined with agentic systems, enabling coordinated AI swarms that amplify manipulation campaigns \cite{schroeder2026malicious}. As Coeckelbergh \cite{coeckelbergh2025llms} notes, the crisis of truth created by LLMs can undermine citizens' capacity for critical reflection, promoting passive consumption over democratic deliberation.

Second, the opacity of LLM outputs and their underlying training data reinforces power asymmetries (AIRD3), as their development is concentrated within a handful of private actors with limited transparency or public oversight. Since these models are often trained on vast corpora of internet data, they may also reproduce social biases and reinforce unfairness (AIRD2) in downstream applications.

Finally, as LLMs are increasingly integrated into civic interfaces, such as chatbots for public services, policy consultations, or legal advice, they raise important questions about democratic accountability, authorship, and the legitimacy of machine-mediated communication in public life.

Despite these challenges, LLMs also offer democratic potential when responsibly designed and deployed. They can support informed citizen participation by simplifying complex legal or policy texts, improving accessibility through multilingual translation, enabling inclusive dialogue on civic platforms, assisting public consultations, and expanding educational outreach on democratic processes. Early work on AI-mediated deliberation also suggests that AI systems may support democratic discussion when they are carefully designed to facilitate mutual understanding rather than replace human judgment \cite{tessler2026mediation}. If transparently governed and aligned with participatory values, LLMs may contribute to a more informed and engaged citizenry.

\subsection{Lessons Learned}

Through the development of the AIRD and AIPD taxonomies, several key insights have emerged regarding the governance of trustworthy AI in democratic contexts:

\begin{itemize}

    \item \textbf{Democracy-specific risks are under-theorized.} Comparative analyses and normative frameworks in AI ethics show that many prominent guidelines converge on a small set of high-level, largely individual-rights based principles  \cite{floridi2019unified,fjeld2020principled,radanliev2025ai}, and say little about how AI systems can threaten democratic values such as collective autonomy, institutional accountability, and pluralism.

    \item \textbf{Trustworthiness must be evaluated relationally.} Trust is not only about system reliability or transparency in isolation; it depends on power relations, perceived legitimacy, and the capacity for democratic oversight.

    \item \textbf{Alignment is not enough.} Value alignment\footnote{Value alignment aims to steer AI behavior toward human values and societal norms. While important, it often focuses on individual aims and can overlook institutional risks such as diminished public reasoning, reduced civic engagement, or concentrated power—issues that require democratic governance and robust institutions beyond technical fixes.} approaches often center on intent and outputs but may ignore structural and institutional risks—such as the erosion of public reasoning or the capture of democratic processes—that cannot be solved through technical fixes alone.

    \item \textbf{Democratic resilience requires adaptive governance.} Left unchecked, AI can reinforce power asymmetries and institutional opacity. Trustworthy AI therefore requires participatory mechanisms, contestability, regulatory responsiveness, and continuing institutional adaptation as technologies, capacities, and societal expectations evolve.

    \item \textbf{Conceptual clarity enables policy relevance.} By distinguishing between risks to individuals and risks to democracy, the framework clarifies where different governance tools (e.g., data protection vs. democratic accountability) are most appropriate.
\end{itemize}

\subsection{Limitations and Future Research Directions}


This study has two important limitations. First, the AIRD/AIPD framework is conceptual and taxonomic. It synthesizes and organizes recurring themes in the literature, but it does not provide empirical validation of the causal effects of specific AI systems on democratic outcomes. Future research should test the framework through comparative case studies, empirical evaluation, expert elicitation, and application across different institutional and cultural contexts.

Second, the worked examples are illustrative applications, not empirical case studies or generalizable evidence. They show how to identify categories, affected democratic principles, Trustworthy AI requirements, governance responses, and responsible actors, while recognizing that outcomes depend on institutional capacity, legal safeguards, deployment context, and public--private power relations.

Applications should also assess impacts across groups. As the public-service example illustrates, allocation systems may affect people who depend on welfare, housing, healthcare, education, and other essential services unequally. Empirical validation should therefore examine how risks and contributions are distributed and whether institutional, technical, and legal governance mechanisms prevent harms from concentrating on disadvantaged communities.

Although not systematically mapped here, trade-offs matter in practice: transparency supports accountability but may conflict with privacy, security, or sensitive information; human oversight strengthens democratic control but may slow decisions and increase administrative burdens; and robustness and explainability improve trust and contestability but require costly expertise, testing, monitoring, and documentation. Risk mitigation therefore also depends on feasibility, institutional capacity, and proportionality.

AI–democracy interactions form a dynamic feedback cycle: deployment produces risks and contributions that prompt contestation, institutional learning, regulation, auditing, and redesign, potentially mitigating harms, redistributing benefits and burdens, or creating new effects. AIRD/AIPD therefore provides a time-sensitive analytical snapshot; longitudinal comparative studies should trace how these interactions and governance responses evolve and reshape democratic institutions.

Finally, we do not provide a full cost, distributional, or capacity analysis of how different democracies could implement the measures associated with our taxonomies. Risk mitigation may require human oversight, documentation, testing, monitoring, auditing capacity, and institutional coordination, which can affect feasibility, especially in low- and middle-income democracies or public institutions with limited regulatory and fiscal capacity.

In addition, the framework is grounded primarily in liberal democratic principles; applying it in hybrid or non-Western contexts may require reinterpretation to support more inclusive AI governance.

Debates on AI personhood, dignity, and governance also warrant study for their effects on regulatory design, trust, and democratic oversight \cite{novelli2024ai,rueda2025dignity,coeckelbergh2025llms}.

Taken together, these directions show that AI--democracy alignment requires empirically informed and evolving governance, not conceptual analysis alone.

\subsection{Responding to Alternative Perspectives}

Critics warn that regulation may stifle innovation, create bureaucratic hurdles, or disproportionately burden smaller developers; techno-optimists instead favor competition, user choice, or open-source transparency. AIRD/AIPD treats autonomy, fairness, accountability, and trust as political goods requiring protection, while recognizing that proportionate governance can support innovation, social trust, and democratic legitimacy.

\section{Conclusion}
\label{sec:Con}

The AIRD/AIPD framework captures AI's dual relationship with democratic governance: AIRD identifies democratic risks, while AIPD identifies conditional contributions. AI is thus neither inherently democratic nor harmful; its effects depend on design, deployment, institutional context, and governance.

The contribution of the framework extends beyond classification. By connecting democratic principles with Trustworthy AI requirements, it provides an operational pathway from democratic diagnosis to governance action: identifying the affected context and actors, classifying primary and secondary AIRD and/or AIPD effects, mapping relevant Trustworthy AI requirements, and defining proportionate governance responses and responsible actors. The operational mapping and worked examples illustrate how this sequence can support both predominantly harmful applications and mixed cases in which democratic contributions coexist with risks.

The framework remains conceptual and has not yet been empirically validated across jurisdictions or application domains. Future research should test its diagnostic consistency, examine real-world case studies, evaluate the effectiveness and costs of governance responses, and investigate how AIRD and AIPD categories evolve as AI systems, regulatory regimes, and democratic institutions change. Particular attention should be paid to resource-constrained democracies, differential impacts on disadvantaged communities, and general-purpose or frontier AI systems whose opacity may limit causal attribution.

Trustworthy AI is an institutional and political condition for democratic resilience. AIRD/AIPD connects democratic diagnosis to responsibility and proportionate intervention. A democratic future with AI requires deliberate, accountable, and democratically legitimate design and governance.

\section*{Acknowledgments}
F. Herrera and O. Mentxaka are supported by Project TSI-100927-2023-1, funded by the European Union's Recovery, Transformation and Resilience Plan (NextGenerationEU) through the Ministry for Digital Transformation and the Civil Service. F. Herrera also received support from the Spanish \emph{Ministry of Science and Innovation} under Project PID2023-150070NB-I00.

\section*{Declaration of competing interest \& Disclaimer}
The authors declare no known competing financial interests or personal relationships that could have influenced this work. The views expressed are solely those of the authors and do not represent an official position of the European Commission.
 
\section*{Credits}
All images were drafted by the authors and illustrated by Pablo García, whose contribution is gratefully acknowledged.

\FloatBarrier
\bibliographystyle{IEEEtran}
\bibliography{refs}

\end{document}